\documentclass[10pt, journal]{IEEEtran}
\pdfoutput=1
\usepackage[pdftex]{graphicx}
\usepackage[T1]{fontenc}% optional T1 font encoding
\usepackage{amsmath}
\usepackage{amsfonts,url}
\usepackage{subcaption}

\newtheorem{theorem}{Theorem}

\newtheorem{proposition}{Proposition}
\newtheorem{definition}{Definition}
\newcommand*{\Rom}[1]{\uppercase\expandafter{\romannumeral #1\relax}} % new command to type upper case Roman numbers
\usepackage{textcomp}
\usepackage{gensymb} % include degree mark
\usepackage{bbm}
\usepackage{authblk}
\usepackage{stackengine}
\usepackage{changes} % used to markup revision: added, deleted, ...
\setaddedmarkup{{\color{orange} #1}} % set markup for added in changes package
\usepackage{multirow} % package for multi row in table
\usepackage{bm}
\usepackage{booktabs} % For formal tables
 \usepackage{amssymb}
 \usepackage{cite}
\newcommand{\brac}[1]{\left(#1\right)}
\newcommand{\sbrac}[1]{\left[#1\right]}
\newcommand{\cbrac}[1]{\left\{#1\right\}}

\newcommand{\bvec}[1]{\mathbf{#1}}
\newcommand{\expect}[1]{\mathbb{E}\sbrac{#1}}

\newcommand{\ceil}[1]{\lceil #1\rceil }

\newcommand{\blockprobclass}[5]{p_{b,#1}\brac{ #2, #3, #4, #5}}

\newtheorem{assumption}{Assumption}

\begin{document}
\title{Resource Allocation and HARQ Optimization for URLLC Traffic in 5G Wireless Networks }

\author{Arjun~Anand,~\IEEEmembership{Student Member,~IEEE,}
        Gustavo~de~Veciana,~\IEEEmembership{Fellow,~IEEE.}\thanks{This  work is supported by Futurewei Technologies.}} 

\maketitle

\begin{abstract}
5G wireless networks are expected to support Ultra Reliable Low Latency Communications (URLLC) traffic which requires very low packet delays ( <  1 msec.) and extremely high reliability ($\sim$99.999\%).  In this paper we focus on the design of  a wireless system supporting downlink URLLC traffic.  Using a  queuing network based model for the wireless system we characterize the effect of various design choices on the maximum URLLC load it can support, including: 1) system parameters such as the bandwidth, link SINR , and QoS requirements; 2) resource allocation schemes in Orthogonal Frequency Division Multiple Access (OFDMA) based systems; and 3) Hybrid Automatic Repeat Request (HARQ) schemes. Key contributions of this paper which are of practical interest  are: 1) study of how the the minimum required system bandwidth to support a given URLLC load scales with associated QoS constraints; 2) characterization of optimal OFDMA resource allocation schemes which maximize  the admissible URLLC load; and 3) optimization of a repetition code based HARQ scheme which approximates Chase HARQ combining.   
\end{abstract}


% \fi
\IEEEpeerreviewmaketitle

\section{Introduction}
%URLLC traffic is an important aspect of 5G networks and provisioning resources is a challenging problem

5G wireless networks are expected to support a new class of traffic called  Ultra Reliable Low Latency Communication (URLLC) for  appplications like industrial automation, mission critical traffic, virtual reality, etc.,  see e.g.,~\cite{hwwtahaa16, ywjbas15, gla17, pbfms16, 3gpp_ran1_87, 3gpp_ran1_88,  Bennis_2018}.  URLLC traffic requires  packet latencies  less than 1 msec along with a very high reliability of 99.999 \%.  Wireless system design to meet such stringent Quality of Service (QoS) requirements is  a particularly challenging task and is the focus of this paper.  Specifically in this paper we  consider downlink transmission of URLLC traffic in a Frequency Division Duplex (FDD) based system with separate frequency bands for uplink and downlink. 

The QoS requirements URLLC traffic places on a wireless downlink system are specified as follows: a packet of  size $L$ bits must be successfully  delivered  to the receiver by the Base Station (BS) within a end-to-end  delay of no more than $d$ seconds with a probability of at least $1-\delta$.  The delay experienced by a packet includes queuing delay at the BS,  transmission duration, receiver processing delay, packet decoding feedback transmission duration, and time to make further re-transmissions as needed. Typical values of QoS parameters mentioned in the literature are $L=32$ bytes, $d=1$ msec., and $\delta=10^{-6}$, see~\cite{3gpp_ran1_88} for more details.  
%The Quality of Service (QoS) requirements URLLC traffic places on the Radio Resource Management (RRM) layer of the protocol stack  are  specified as follows: A packet of  size $L$ bits must be successfuly  delivered  to the receiver by the Base Station (BS) within a end-to-end  delay of no more than $d$ seconds with a probability of at least $1-\delta$.  The delay experienced by a packet includes queuing delay at the BS,  transmission duration, receiver processing delay, packet decoding feedback transmission duration, and time to make further re-transmissions as needed. Typical values of QoS parameters mentioned in the literature are $L=50$ bytes, $d=1$ msec, and $\delta=10^{-6}$, see~\cite{3gpp_ran1_88} for more details.  

 This paper investigates how design choices impact the  URLLC `capacity',  i.e., the maximum URLLC load the system can support and how this is affected by the stringency of the QoS requirements. In particular, the paper  studies the impact of: 1) system bandwidth $W$, user SINR, QoS parameters $d$ and $\delta$;  2) resource allocation in the time-frequency plane of an Orthogonal Frequency Division Multiple Access (OFDMA) system; and 3), Hybrid Automatic Repeat Request (HARQ) scheme on URLLC capacity. The three aspects are inter-related, for example, the impact of the system bandwidth $W$ on URLLC capacity depends on the HARQ schemes being used. 
 
  Our aim of understanding of the impact of the system parameters on URLLC capacity will help system engineering to meet URLLC's QoS requirements. Another important aspect which needs careful consideration is how resources are  allocated to  URLLC transmissions. 5G standards are  OFDMA based and hence, users' packets are allocated different parts of a  time-frequency plane for data transmission. To send a URLLC packet, one can schedule  `tall' transmissions which use a  large swath of bandwidth for a short duration or  `wide' transmissions which use a small bandwidth over a longer duration.  `Tall' transmissions result in reduced transmission times for packets, however, the maximum number of concurrent transmissions is also reduced. This might result in queuing or blocking of URLLC packets due to the immediate un-availability of bandwidth.  By contrast, `wide' transmissions  permit a higher number of concurrent transmissions but with  longer transmission times for each packet. Hence, the average number of  packets in the system  is higher with `wide' transmissions than with `tall' transmissions, which may again lead to bandwidth scarcity. Further the transmission duration for `wide' transmissions are constrained by $d$. Therefore, one would like to analyze the trade-off between `tall' and `wide' packet transmissions. 
%  
%   A characterization of the impact of resource allocation scheme can help  one decide whether one should schedule `tall' transmissions which use a  large bandwidth for a short duration or  `wide' transmissions which use a small bandwidth over a longer duration.

Finally studying the impact of HARQ schemes on URLLC capacity can help evaluate various design choices, such as  the maximum number of re-transmissions allowed and the reliability (coding scheme) one should target after each transmission. This paper proposes an analytical framework to study the above mentioned design choices and trade-offs.

\subsection{Related Work}
Questions surrounding URLLC traffic have recently received a lot of attention. The 3GPP standards committee has recognized the need for a new  OFDMA based frame structure to support such traffic, see e.g.,~\cite{3gpp_ran1_88} for a discussion of various proposals. In particular to meet the stringent latency constraints of URLLC traffic, they have proposed a \emph{mini-slot}  level access to  radio resources for URLLC traffic with  mini-slot durations of $0.125-0.250$ msec. This is different from the standard \emph{slot} level access to radio resources for eMBB traffic which has slot  durations of $1$ msec or higher. 
%The use of flexible traffic dependent slot durations and resource allocation has also been proposed in~\cite{pbfms16}. 

System level designs for URLLC networks have been explored in~\cite{ Bennis_2018, Popovski_2017,Ji_2017,Ashraf_2017,ljcjs17, dkp16}.  In~\cite{ Bennis_2018}, the authors have surveyed the literature on URLLC traffic and have elaborated on the various technologies and methodologies related to URLLC system design.   In~\cite{dkp16}, the authors discuss  information theoretic results on sending short packets. They also discuss protocols to transmit small length packets between two nodes in a downlink broadcast setting and for random access based uplink. However, they do not focus on optimizing the resources required in an OFDMA based system supporting stochastic loads.   In~\cite{Popovski_2017}, the authors have covered various aspects of URLLC traffic like the overhead due to packet headers, decoding failure probability of URLLC transmissions, and Channel State Information (CSI) acquisition at the transmitter. 
%They have also proposed using interface diversity, grant-free access for uplink and device-to-device communication (D2D) as possible solutions to achieve the stringent latency requirements of URLLC traffic.
 In~\cite{Ji_2017}, the authors discuss  QoS requirements for URLLC traffic. They also specify various methods to share resources among URLLC and other traffic types. In~\cite{Ashraf_2017}, the authors study the effect of physical layer waveforms, OFDMA numerology, and Forward Error Correction (FEC) schemes on  URLLC capacity via simulation. They further propose the use of Tail Biting Convolution Codes (TBCC) to achieve a target reliability as high as $10^{-9}$.

The work in~\cite{ljcjs17} is most closely related to ours. The authors have  used a queue based model and simulations to study the  design of wireless systems supporting  URLLC traffic. In particular they introduce simple $M/M/m/k$ and $M/D/m/m$ queuing models to study trade-offs among system capacity, latency  and reliability requirements for the worst case scenario where  all users are at the cell edge.      In particular, they have  considered  trade-offs among system capacity, reliability, and latency requirements.     However, in the analysis of system trade-offs, they have only considered packet loss due to `blocking' at the BS, i.e., unavailability of resources to immediately transmit a packet, and have not explicitly considered the effect of decoding failures and re-transmissions on system capacity.  Also, they do not consider the optimization of  HARQ schemes. Our work is inspired by this initial work's approach.  

 The above mentioned work~\cite{ljcjs17}  also focussed on multiplexing of enhanced Mobile Broadband (eMBB) and  URLLC traffic. They showed that allocating dedicated frequency bands to URLLC and eMBB traffic is inefficient, and have advocated a shared wide-band resource allocation for both URLLC and eMBB traffic. In addition to~\cite{ljcjs17}, there are a few other works~\cite{Anand_2017 , You_2018} which address multiplexing URLLC and eMBB traffic via preemptive puncturing/superposition of eMBB traffic. 
% 
% In~\cite{Anand_2017}, we have also considered  the multiplexing of eMBB and URLLC traffic via puncturing/superposition of eMBB traffic and developed joint scheduling policies for eMBB and URLLC. 

The line of work~\cite{Shariatmadari_2015, Shariatmadari_2016} on HARQ design and optimization for URLLC traffic is closely related to our work. However, the key difference is that they focus only on the mean resource utilization of various HARQ schemes, whereas we focus on both the \emph{mean} and \emph{variance} of resource utilization of HARQ schemes in an OFDMA based system.  This leads to solutions which are different from the ones obtained by minimizing just the mean resource utilization. 

Another work which is related to ours is~\cite{Malak_2017}. In~\cite{Malak_2017}, the authors have considered the design of random access strategies for uplink delay sensitive communications. In particular they have optimized the number of frequency bins and  HARQ stages under various SINR regimes for chase HARQ combining. Our focus in this paper is on scheduled downlink communications which is different from their system model.  

Many works focus on the industrial applications of URLLC traffic and exhibit simulation based studies for such  systems, see~\cite{gla17,ywjbas15,hwwtahaa16}.  Some  authors, see e.g.,~\cite{dkopy16, sltum16} focus exclusively on physical layer aspects like modulation and coding,  fading and link budget analysis.   However, the above mentioned works do not holistically address  the  design of wireless systems supporting URLLC traffic.

% To study the capacity URLLC packet transmissions which are generally, we have used the channel capacity results under the finite block length regime elaborated in~\cite{} in our analysis. We have also used some of the classical results from queuing theory (~\cite{}) and loss networks in our analysis of URLLC capacity.   

%Compare and contrast with Qualcomm paper. 

\subsection{Our Contributions}
In this paper we shall consider a simple Poisson model for URLLC packet arrivals. In line with the previous works, we shall also assume a wide-band allocation of resources to  URLLC traffic by considering systems where such traffic can preemptively puncture/superpose URLLC packets upon previously scheduled eMBB traffic when necessary. We thus assume URLLC packet transmissions are scheduled immediately upon arrival.  Such a model is not unreasonable due the stringent latency requirements of URLLC traffic. Based on this model the paper makes the following key contributions.
\begin{enumerate}
\item {\em  Resource allocation in OFDMA systems: } We initially consider a \emph{one-shot transmission} model (re-transmissions not permitted) we show that extending URLLC transmissions in time (while reducing the corresponding bandwidth usage) subject to deadline constraints increases the URLLC load that can be supported, i.e.,  `wide' transmissions are better than `tall' transmissions. 
\item {\em Impact of system parameters: } Using an extension of the classical square-root staffing rule, we characterize the minimum overall system bandwidth $W$ needed to support a given URLLC load. Further using the channel capacity results of~\cite{Polyanskiy_2010} in the finite blocklength regime we study the scaling of URLLC capacity as a function of  $W$, SINR, $d$ and $\delta$.    
\item {\em Modeling re-transmissions/HARQ:}  We extend the one-shot transmission model to incorporate  HARQ schemes which  allow re-transmissions if needed.  The entire downlink system, the BS and associated users are modeled  as a  queuing network.  In this setting we  derive closed form  expressions for various important parameters of the system such as average packet delay, distribution of the number of packets in the system, average bandwidth utilization, etc.  Our framework allows us to explore the effect of a given  HARQ scheme on the URLLC capacity.  
\item {\em Optimization of HARQ:}  Finally we consider the optimization of HARQ schemes to minimize the necessary bandwidth required to support a given URLLC load. This can be viewed as  the dual of URLLC capacity maximization problem. We consider a repetition coding scheme which can approximate \emph{Chase combining} for URLLC traffic. We identify two relevant regimes of operation, namely,  \emph{variance dominated} (for overall low URLLC loads) and \emph{mean utilization dominated} ( for high overall URLLC loads) regimes, and  reach the following two conclusions:

a) In the variance dominated regime, a one-shot transmissions are optimal. 
%a) At low loads, the required system bandwidth $W$ is minimized when we use only one transmission with appropriate coding to meet the reliability requirement  while spreading out the transmission in time as much as possible without violating the deadline d. This holds for a range of SINRs and packet sizes.

b) In the mean utilization dominated regime, the optimal transmission scheme leverages re-transmissions, if necessary, to meet the desired reliability requirement. Further the maximum number of re-transmissions permitted in the optimal scheme is a non-increasing function of the SINR. 
\end{enumerate}    
\subsection{Organization}
The paper is organized as follows. In Sec.~\ref{sec:one_shot_transmission} we introduce our one-shot transmission model and develop the key associated results. In Sec.~\ref{sec:retransmissions} we extend the one-shot transmission model to incorporate  HARQ schemes. In Sec.~\ref{sec:urllc_capacity} we discuss the optimization of  HARQ schemes to maximize URLLC capacity. This is  followed by  conclusions in Sec.~\ref{sec:conclusions}.

\section{Performance Analysis: One-Shot Transmission}
\label{sec:one_shot_transmission}
%We consider downlink scheduling of URLLC traffic where such traffic requires a maximum
%packet transmission delay  less than $d$ seconds and extremely high reliability, e.g., a probability of failure $ \delta $ no more than $ 10^{-6} $.
 In this paper we focus on  downlink transmissions in a wireless system with a single Base Station serving a dynamic population of  URLLC users and their associated packets.  The wireless system is OFDMA  based where different parts of the time-frequency plane are allocated to URLLC users' packets  based on  transmission requests. A URLLC packet may suffer from queuing delays at the BS, transmission and propagation delays, and receiver processing delays. The system should be engineered such that the QoS requirements of URLLC traffic are satisfied, i.e., a URLLC packet of size $L$ bits  must be delivered successfully to the receiver within a total delay of $d$ seconds with  a success probability of at least $1-\delta$. We  start by introducing our system model.
 \subsection{System Model-- One Shot Transmission} 
  We consider a system operating in a large aggregate bandwidth of say $ W $ Hz\footnote{This need not be a contiguous bandwidth, but result from the use of carrier aggregation across disjoint segments}. For simplicity we ignore the slotted nature of the system. To model the `near far' effects in  wireless systems, we shall consider a multi-class system with $C$ classes of users where each class represents users with same SINR\footnote{Ideally SINR is a continuous random variable, however, in practical systems the channel quality feedback from users are quantized to several discrete levels. }. The aggregate traffic generated at the BS by class $c$ users is modeled  as a  Poisson process with rate $ \lambda_c $ packets/sec. Define the vector of arrival rates $\boldsymbol{\lambda}:= \brac{\lambda_1, \lambda_2, \ldots, \lambda_C}$. Let $SINR_c$ denote the  SINR of a class $c$ user's packets. 
  
  We  initially assume that each URLLC packet is transmitted once. We will call this the \emph{one-shot } transmission model.  We will extend this to include re-transmissions in Sec.~\ref{sec:retransmissions}.  A packet destined to a class $c$ user requires $r_c$ channel uses in the time-frequency plane to transmit its codeword. The codeword for a transmission is  chosen such that the decoding is successful with probability of at least $1-\delta$.  A  URLLC packet of class $c$ is allocated a bandwidth of $ h_{c} $ for a period of time  $ s_{c} $. These values are fixed and related to $r_c$  by $  \kappa s_{c}h_{c} =  r_{c}$, where $\kappa$ is a constant which denotes the number of channel uses per unit time per unit bandwidth of the OFDMA time-frequency plane. The value of $\kappa$ depends on the OFDMA frame structure and numerology.   Since URLLC packets have a deadline of $d$ seconds, we shall always choose $s_c \leq d$. For  ease of analysis we shall also assume that for any class $c$,  $d$ is an integer multiple of $s_c$.  Thus following vectors which characterize the system: $\bvec{r}:= \brac{r_1, r_2, \ldots, r_C}$, $\bvec{s}:=\brac{s_1, s_2, \ldots, s_C}$,   $\bvec{h} := \brac{h_1, h_2, \ldots, h_C}$ and  $\boldsymbol{\rho}:= \brac{\rho_1, \rho_2, \ldots, \rho_C}$, where $\rho_c:=\lambda_c s_c$.

  We shall make the following key assumption on the system operation.
\begin{assumption} 
\label{asm:assumptions_on_multi_class_model_one_shot}
%\vspace{-0.2in}  
 ({ \bf Immediate scheduling }) A  URLLC packet transmission request is scheduled immediately upon arrival if there is spare bandwidth is available. Otherwise the packet is lost. New packets do not preempt ongoing URLLC packet transmissions.  
\end{assumption}
%{\em Remarks:} We  model  heterogeneous wireless channel conditions of users in a wireless system via a multi-class  system. Each class is a set of  users sharing similar channel quality. We assume that a user does not change its class while the BS is in the process of delivering a URLLC packet. This is a reasonable assumption because the deadline $d$ to deliver the packet is typically less than  a millisecond. In this section we shall assume a one-shot transmission model where a packet is transmitted once with no re-transmissions. In line with the previous work, we shall also consider a wideband resource allocation model of URLLC traffic with the entire bandwidth $W$ available for  URLLC traffic.

 Given the stringent latency requirements, the immediate scheduling assumption is a reasonable design choice. 
% Due to the immediate scheduling assumption  URLLC traffic  preemptively punctures/superposes previously scheduled eMBB traffic in the time and frequency plane. 
 % Further our wideband resource allocation model ensures that queuing event may occur very rarely..   
%No preemption assumption is justified because the system would be engineered such that the blocking events are rare and the  allowing  preemption or not would not make much difference in the system capacity analysis in the sequel. 
\subsection{Infinite System  Bandwidth }
Initially let us consider a system with infinite bandwidth, i.e., $W=\infty$. In such a system the base station can be modeled as a multi-class $M/GI/\infty$, see~\cite{kleinrockI75_book} for more details.
% The packet arrival process of any class is a Poisson process and hence, it is denoted by the letter $M$ which stands for memoryless arrivals in the Kendall's notation for queuing systems~\cite{kleinrockI75_book}.  The service times for packet transmissions in idfferent transmissions are given by  $\bvec{s}$ are different across classes but are  homogeneous and determinstic for each class. Therefore, it is denoted by the term $GI$ which stands for general and independent service times for packets.  The  $\infty$ in this notation stands for the number of parallel transmissions that can be supported which is infinite because of our initial  assumption of infinite bandwidth. 
Let $\bvec{N}:= \brac{N_1, N_2, \ldots, N_C}$ be a  random vector denoting the number of active transmissions when the system is in steady state. For any $\bvec{n} \in \mathbb{Z}_+^{C}$, let $\pi(\bvec{n}):= P(\bvec{N}=\bvec{n})$ be the stationary distribution. Using  standard results for $M/GI/\infty$ queues (see~\cite{Harchol-Balter_book}) one immediately gets the following results: 
\begin{equation}
\label{eq:distribution_one_shot}
\pi\brac{\bvec{n}} = \Pi_{c=1}^{C}  \brac{ \frac{\rho_{c}^{n_{c}}}{n_{c} !}} \exp\brac{-\rho_{c}},
\end{equation}
 and the average bandwidth utilization is given by $$\expect{\bvec{h}\bvec{N}^{T}}=\bvec{h}\boldsymbol{\rho}^{T}.$$
  Observe that the number of active transmissions  of any class $c$ is  Poisson distributed with mean $\rho_c$.   Thus  $\rho_c$ as the average load of  class $c$ traffic. 
%\begin{figure*}
% \centering
%  \includegraphics[width=0.55\textwidth]{one_shot_transmission}
%  \caption{Base Station in a multi-class wireless system with $W=\infty$  modeled as an $M/GI/\infty$ queue. In this example $C=2$ and $\bvec{h}$ and $\bvec{s}$ are such that $h_1 < h_2$ and $s_1 > s_2$. }
%  \label{fig:one_shot_queuing_system}
%\end{figure*} 
       
\subsection{Effect of Finite System Bandwidth}
Although in practice the available system bandwidth $W$ is not infinite but possibly large.  We will consider a case where   a wide-band  allocation $W$ is available  to transmit URLLC traffic. This might be made available through a puncturing/superposition scheme between URLLC and eMBB traffic. see e.g.,~\cite{Anand_2017}.  Even  large bandwidth systems can occasionally suffer from congestion due to the stochastic variations in the arrival process and occasionally there may not be enough spare bandwidth to transmit a new URLLC packet. In such cases we shall assume that  packets are \emph{blocked} and dropped from the system.    Let $\bvec{N}(t):=\brac{N_1(t), N_2(t), \ldots, N_C(t)}$ be a random vector denoting the number of   packets of each class in the system at time $t$.   A  class $c$ packet arriving at time $t$ is blocked if the following condition holds:
\begin{equation}
\label{eq:blocking_condition}
h_c + \sum_{c'=1}^{C}h_{c'} N_{c'}(t) > W.
\end{equation}
%Blocking arriving packets based on~\eqref{eq:blocking_condition} also ensures that an ongoing transmission is never preempted by a new packet. 
%In addition to a  blocking event, there are two other means by which packets may fail. First,  a transmission may fail due to a decoding failure at the receiver. We have already assumed that $r_c$ is chosen such that the probability of decoding failure is of the order of $\delta$. Secondly, the transmission duration may exceed the deadline $d$. Note that because of the deterministic service time for a packet, either all transmissions of a class meet the deadline or violate the deadline with probability one. Therefore, we restrict ourselves to the case when deadline is met with probability one, i.e, for  all $c$, $s_c \leq d$. 
We address the following two questions in this section:
\begin{enumerate}
\item How do the choices of  $\bvec{h}$ and $\bvec{s}$ affect the blocking of URLLC packets?
\item What is the required  system bandwidth $W$ given a  desired packet reliability $\delta$? 
\end{enumerate}

To study the effect of $\bvec{h}$ and $\bvec{s}$ on the blocking of URLLC traffic, we shall first consider the blocking probability of a typical class $c$ packet. Observe that the  blocking probability experienced by packets of a class depends on $\bvec{h}, \bvec{s}$ (of all classes), $\boldsymbol{\lambda}$ and $W$. 
%Define $\bvec{h}:=\sbrac{h_1, h_2, \ldots, h_C}$ and $\bvec{s}:= \sbrac{s_1, s_2, \ldots, s_C}$. 
Let $\blockprobclass{c}{\bvec{h}}{\bvec{s}}{\boldsymbol{\lambda}}{W}$ be the blocking probability experienced by a typical class $c$ packet arrival. The fraction of class $c$ traffic admitted is then given by $\lambda_c \brac{1-\blockprobclass{c}{\bvec{h}}{\bvec{s}}{\boldsymbol{\lambda}}{W}}$. Hence, lowering the blocking probability increases the admitted URLLC traffic.     The following result which is proved in Appendix\ref{pf:multi_class_block_prob} gives us the key insight on  optimal choices of $\bvec{h}$ and $\bvec{s}$ for URLLC packet transmissions. 
\begin{theorem}
\label{thm:multi_class_block_prob}
For a given $\bvec{h}$ and $\bvec{s}$,  positive integer $q$,  and  $i\in \cbrac{1, 2, \ldots, C}$ define $\bvec{h}':=\brac{h_1, h_2, \ldots, h_i/q, \ldots, h_C}$ and  $\bvec{s}':=\brac{s_1, s_2\ldots, qs_i, \ldots, s_C}$. Under the one-shot transmission model and Assumption~\ref{asm:assumptions_on_multi_class_model_one_shot}, if $ \rho_i < 1$, then    for any  $c \in \cbrac{1,2, \ldots, C}$, there exists $\tilde{W}_c$ such that for $W > \tilde{W}_c$ we have that $\blockprobclass{c}{\bvec{h}}{\bvec{s}}{\boldsymbol{\lambda}}{W} \geq  \blockprobclass{c}{\bvec{h}'}{\bvec{s}'}{\boldsymbol{\lambda}}{W}$. 
\end{theorem}
{\em Remarks:}
Observe that in wide-band systems scaling $h_i$ and $s_i$ by an integer $q$ as required in the above theorem  increases the number of concurrent transmissions of class $i$ and is also beneficial for all classes (including class $i$). To understand this let us look at the mean and variance of the bandwidth utilization of class $i$ in a  system with parameters $\bvec{h}'$ and $\bvec{s}'$ and infinite bandwidth. The average bandwidth utilization of class $i$, given by $h_i \lambda_i s_i$, does not change with scaling factor $q$, as the decrease in bandwidth of class $i$ is compensated by corresponding increase in the average number of users of class $i$. However, the variance of the bandwidth utilization, given by $\frac{1}{q} h_i^2 \lambda_i s_i$ decreases with $q$.  Therefore, the  congestion events occur less frequently and the system admits more traffic.  Note that this observation is in line with the previous work on URLLC traffic (see~\cite{Bennis_2018}) where the emphasis is on such events corresponding to  the `tail' of URLLC traffic demand.  Further, the assumption $\rho_i < 1$ is not restrictive as one can divide a class into various `virtual' sub-classes such that the average load in each sub-class is less than unity.  

%In the statement of the theorem we require the wide-band condition, i.e., $W > \tilde{W}_c$ because the quantization effects in the maximum number of concurrent transmissions that can be admitted is negligible.  Since, we require very high reliability for URLLC traffic (extremely low blocking probability),  such a condition is not unreasonable.  

% Note that the above result will not hold if the scaling is not done by an integer $q$. 
%This suggests that if one decreases the bandwidth allocation $h_c$ or equivalently spreads the transmission time $s_c$ such that the  number of concurrent transmissions of class $c$ increases, then the blocking  probability decreases. 

Therefore, one should scale $s_i $ with an integer $q$ such that $qs_i=d$.  Such an integer $q$ exists because of our assumption that $d$ is an integer multiple of $s_i$. Hence,  this motivates the following optimal choices of $s_i$ and $h_i$: 
\begin{equation}
s_i=d \quad \text{and} \quad h_i= \frac{r_i}{\kappa d}. 
\end{equation} 
%However, it may not be possible to find and an integer $q$ which meets equality $qs_c=d$ due to quantization effects.  Because of our assumption of wide-band resource allocation for URLLC traffic, the quantization efffects .     
%
%quantization effects that would preclude full utilization of the
%bandwidth by a maximal number of concurrent transmissions, in terms of blocking probability,
To summarize, one might think that `tall' transmissions are better as they take  less time, however, according to the above result  it is better to decrease the bandwidth per transmission and spread out the transmissions as `wide' as possible in the time axis, i.e.,  increase  $s_{i}$ (and decrease $h_{i}$) as long as the deadline is not violated.  

 To meet the reliability requirements of URLLC traffic, the system  bandwidth $W$ must be chosen such that the probability of blocking of a typical URLLC packet arrival is of the order of $\delta$. To that end we shall use a multi-class extension of the classical square-root staffing rule (see~\cite{Harchol-Balter_book} for more details) to relate $W$, $\bvec{r}$, $\boldsymbol{\lambda}$ and $\delta$.  Under this dimensioning rule, to support a URLLC load of $\boldsymbol{\lambda}$ with reliability  $\delta$ for a given $\bvec{r}$, the system bandwidth should satisfy the following condition:
 \begin{equation}
\label{eq:multi_class_sqrt_staffing_rule_one_shot}
    W \geq \zeta^{\text{mean}}\brac{\bvec{r}} + c(\delta) \sqrt{\zeta^{\text{variance}} \brac{\bvec{r}}}, 
 \end{equation}  
where $c(\delta)= Q^{-1}\brac{\delta}$, $Q\brac{\cdot}$ is the Q-function,   ${\zeta}^{\text{mean}}\brac{\bvec{r}}:= \sum_{c=1}^{C}  \lambda_{c} \frac{r_{c}}{\kappa} $ is the mean bandwidth utilization, and  ${\zeta}^{\text{variance}}\brac{\bvec{r}}:= \sum_{c=1}^{C}  \lambda_{c}  \frac{r_c^2}{\kappa^2 d}$ is the variance of the bandwidth utilization.    

Next we  study the  URLLC capacity scaling with respect to $W$,  $SINR_c$, $d$, and $\delta$. This  requires a model relating $r_c$, $SINR_c$, and $\delta$ which  is described in the next subsection.  
\subsection{Finite Block Length Model}
\label{subsec:fbl}
 Since the URLLC packet sizes are typically  small, we shall use the capacity results for the finite blocklength regime developed in~\cite{Polyanskiy_2010}.  In an AWGN channel the number of information bits $L$ that can be transmitted  with a codeword decoding error probability of $p$ in $r$ channel uses is given by 
\begin{equation}
\label{eq:block_length_most_exact}
  L =  r C(SINR_c)  -  Q^{-1} \brac{p} \sqrt{r V(SINR_c)} + 0.5 \log_{2}\brac{r} + o(1),
 \end{equation}  
  where $C(SINR_c)= \log_2 \brac{1+SINR_c}$ is the AWGN channel capacity under infinite blocklength assumption and $V(SINR_c)= \brac{\log_{2}(e)}^2\brac{1- \frac{1}{\brac{1+ SINR_c}^2}}$.   Using the above model one can  approximate $r$ as a function of $p$ as follows:
\begin{multline}
\label{eq:block_length_approximation}
r \approx \frac{L}{C(SINR_c)}  + \frac{\brac{Q^{-1}\brac{p}}^2V(SINR_c)}{2\brac{C(SINR_c)}^2}  \\ + \frac{\brac{Q^{-1}\brac{p}}^2V(SINR_c)}{2\brac{C(SINR_c)}^2} \sqrt{1 + \frac{4LC(SINR_c)}{V(SINR_c) \brac{Q^{-1}\brac{p}}^2 }}.
\end{multline}  
A derivation of this approximation is given in Appendix\ref{pf:approx_expression_for_r}. We can now write $r_{c}$ as a function of $\delta$, $L$ and $SINR_c$  for various user/packet classes.   
\subsection{Capacity Scaling}
%We shall first study the scaling result of URLLC capacity for one-shot transmission. 
%
% For simplicity we shall make the following assumption on the arrival rates. 
%\begin{assumption}
%The arrival rates of all classes are equal, i.e.,  for all $c$, $\lambda_c=\lambda$. 
%\end{assumption}
We shall define  the \emph{single class URLLC capacity}  as follows. 
\begin{definition}
For any class $c$, its single class URLLC capacity $\lambda_c^*$ is the maximum URLLC arrival rate that can be supported by the system while satisfying the QoS requirements  if only class $c$ traffic is present in the system. 
\end{definition}

 Note that $\lambda_c^*$ is a function of $W$, $d$. $\delta$, $SINR_c$, and $L$. We would like to study  the scaling of $\lambda_c^*$ with respect to various system parameters. Recall that for $f,g: \, \mathbb{R}_+ \to \mathbb{R}_+$, we say that $f(x)\sim \Theta\brac{g(x)}$ if there exist $x_o$,  $a$, and $b$ such that $a \leq b$ and for $x \geq x_o$ we have that
 \begin{equation}
 ag(x) \leq f(x) \leq bg(x). 
 \end{equation}

 The following result summarizes the scaling of $\lambda_c^*$ with various system parameters.  The proof of the theorem below is given in Appendix\ref{pf:scaling_result}. 
\begin{theorem}
\label{thm:scaling_result}
 Under one-shot transmission model and Assumption~\ref{asm:assumptions_on_multi_class_model_one_shot}  we have that
 \begin{enumerate}
 \item  $\lambda_c^* \sim \Theta \brac{W - \sqrt{W}}$.
 \item For $SINR_c \gg 1$, we have that  $\lambda_c^* \sim \Theta \brac{\log_2 \brac{SINR_c}  - \sqrt{ \log_2 \brac{SINR_c}}}$. 
 \item $\lambda_c^* \sim \Theta \brac{1  - \frac{1}{\sqrt{d}}}$.
 \item $\lambda_c^* \sim \Theta \brac{\frac{1}{-\log_2(\delta)}}$.  
\end{enumerate}  
\end{theorem} 
{\em Remarks:}  Observe that $\lambda_c^*$ scales as a strictly concave function of $SINR_c$, $d$, and $\delta$.  Hence, while increasing $SINR_c$ and $d$ or decreasing $\delta$ one suffers from  diminishing returns. However, as expected the scaling of $\lambda_c$ with respect to $W$ does not suffer from diminishing returns.  For large $W$,  $\lambda_c^*$ increases linearly with $W$ which  is the best one could hope.  
\section{Performance Analysis with Multiple  Transmissions}
\label{sec:retransmissions}
Next we  extend the system model to include re-transmissions and study the role of HARQ schemes.  We shall first explain the extension of our system model. 
\subsection{System Model-- Multiple Transmissions}
Paralleling the one-shot transmission model considered in Sec.~\ref{sec:one_shot_transmission}, we shall consider a multi-class system with Poisson arrivals for URLLC traffic, where a class represents users' packets sharing the same SINR . However, by contrast to our one-shot transmission model, in this section we shall permit packet re-transmissions. Suppose a class $c$ packet can have up to $ m_c $ transmission attempts after which it is dropped. We index transmission attempts by $m=1,2, \ldots, m_c$, where $m=1$ corresponds to the \emph{initial} transmission and any $m>1$ corresponds to a \emph{re-transmission}. A class $c$ packet in the $m^{\text{th}}$ transmission attempt is assumed to require $r_{c,m}$ channel uses in the time-frequency plane. The bandwidth used and the time to transmit in the $m^{th}$ packet transmission are denoted by $h_{c,m}$  and  $s_{c,m}$, respectively.  They are related to $r_{c,m}$ by $\kappa h_{c,m}s_{c,m}=r_{c,m}$.  For any $m \in \cbrac{1,2, \ldots, m_c}$, define $\bvec{r}_c^{(m)}:=\brac{r_{c,1},r_{c,2},\ldots, r_{c,m}}$.    After every transmission  the intended receiver sends a one bit feedback to the BS indicating success/failure of the packet decoding process.   In general, the probability of decoding failure  of a class $c$ packet after the $m^{\text{th}}$ transmission  attempt, denoted by $p_{c,m} \brac{\bvec{r}_c^{(m)}}$,  is a function of $\bvec{r}_c^{(m)} $.  A  decoding failure for a class $c$ packet occurs if the packet has not been successfully decoded after $m_c$ transmission attempts. This happens with probability $\Pi_{m=1}^{m_c} p_{c,m}\brac{\bvec{r}_c^{(m)}}$.   Thus one should design  the system such that $\Pi_{m=1}^{m_c} p_{c,m}\brac{\bvec{r}_c^{(m)}} \leq \delta$. Therefore, the values of $r_{c,m}$, $p_{c,m}\brac{\bvec{r}_c^{(m)}}$ and $m_c$ jointly characterize  the HARQ scheme used for class $c$ users.

  The feedback on success/failure of a transmission will incur   propagation delays, receiver processing delay, and the uplink channel access and scheduling delays. We shall assume that the uplink channel is well provisioned so that there are no scheduling and channel access delays.   Therefore, the total feedback delay  includes only the propagation delay and the receiver processing delay which  we shall denote  by a deterministic value $f_c$ for a class $c$ user. A class dependent feedback delay is consistent with our notion that  classes denote  users with similar channel characteristics, for example, users at the cell edge may experience longer feedback delays.

For any class $c$, define the following vectors
$\bvec{s}_c := \brac{s_{c,1}, s_{c,2} \ldots, s_{c,m_c}}$,  $\bvec{h}_c :=\brac{h_{c,1}, h_{c,2} \ldots, h_{c,m_c}},$  and $\boldsymbol{\rho}_c := \brac{\rho_{c,1}, \rho_{c,2} \ldots, \rho_{c,m_c}}$, 
where $\rho_{c,1}:= \lambda_c s_{c,1}$ and for any $m >1 $ let $\rho_{c,m}:= \lambda_{c} \brac{\Pi_{k=1}^{m-1} p_{c,k}\brac{\bvec{r}_c^{(k)}}}  s_{c,m}$. Using the above definitions, we further define the following vectors capturing the overall system's design and loads: $\bvec{s} := \brac{\bvec{s}_{1}, \bvec{s}_{2}, \ldots, \bvec{s}_{C}}$,   $\bvec{h} :=\brac{\bvec{h}_{1}, \bvec{h}_{2}, \ldots, \bvec{h}_{C}}$,  and  $\boldsymbol{\rho} := \brac{\boldsymbol{\rho}_1, \boldsymbol{\rho}_2, \ldots, \boldsymbol{\rho}_C}$. 
We further let $\bvec{m} := \brac{m_1, m_2, \ldots, m_C}$ denote vector of maximum transmission attempts per class. 

We shall also revise the \emph{immediate scheduling} assumption for the setting with packet re-transmissions. 
%We shall substitute  the assumptions on \emph{one-shot transmission}, \emph{heterogeneous resource requirement} and \emph{resource allocation} in Assumption~\ref{asm:assumptions_on_multi_class_model_one_shot} with the following assumptions, respectively.
\begin{assumption} 
\label{asm:assumptions_on_multi_class_model}
%~\\
%\texttt{\vspace{-0.2 in}
%\begin{enumerate}
%	\item {\em Multi-class system: } There are $C$ user classes  and the URLLC packets destined  to class $c$ users arrive at the BS at a rate of $ \lambda_c $ packets/sec.
%	\item {\em Uniform QoS requirements:} A packet of any class must be decoded successfully by the receiver within a total delay of $d$ seconds with  a success probability of $1-\delta$. 
%\item {\em Retransmissions allowed:}  Each URLLC packet of a class $c$ can have upto    
%	\item {\em Heterogeneous resource requirements:} We assume  each URLLC packet transmission  has to reliably transmit  a packet of size $ L $ bits. We shall assume that packets destined to a class $c$ user require $r_c$ resource elements in the time-frequency plane. 
%	 We shall assume that receiver utilizes chase combining to combine successive re-transmissions to decode     
%	\item {\em Non-idling admission control:} A new URL}LC is admitted into the system if there is spare bandwidth to accommodate it.
%	%	\item {\em Non-anticipative admission control:} A URLLC request admission policy is said to be non-anticipative if it does not make its decision based on the realizations of the r.v. $R$, instead knows only the mean $ \overline{r}$.
%	\item{\em Bandwidth allocation: } A  URLLC packet of class $c$ is allocated a bandwidth of $ h_{c,m} $ Hz in the $m^{\text{th}}$ transmission attempt. Let $ s_{c,m} $ be the time spent by a transmission of class $c$. Its value is deterministic and is given by $ s_{c,m}=r_{c,m}/h_{c,m} $.  
%%	\item {\em Probability of success in a stage:}    
 ({\bf Immediate scheduling  }) An initial URLLC packet transmission request or a   re-transmission is admitted and scheduled for transmission immediately if there is spare bandwidth available. Otherwise the packet is lost. 
%	\item {\em Feedback delay:} For each transmission from the BS the intended receiver sends a feedback through an appropriate uplink channel indicating success/failure of the packet decoding process. Let $f_c$ be the total delay of the feedback process for a class $c$ user. This includes the propagation delays in both the directions, receiver processing time, the time to access the uplink channel and the transmission time of the feedback signal.    
%	\item {\em No preemption: } An ongoing URLLC transmission is never preempted by a new transmission  request. 
%\end{enumerate}
\end{assumption}
\begin{figure*}
 \centering
  \includegraphics[width=0.85\linewidth]{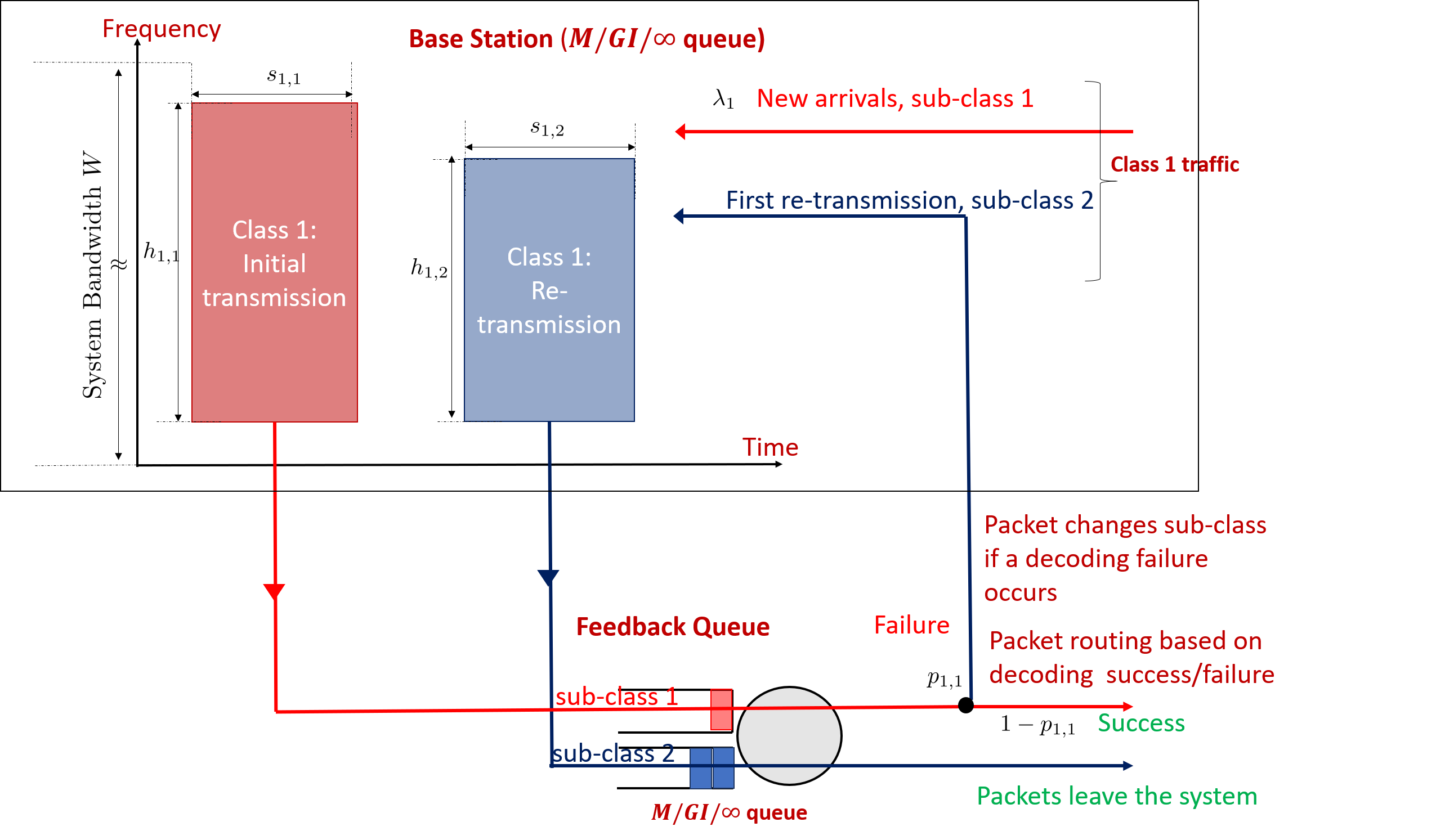}
  \caption{A  wireless system with  a single class of URLLC users modeled as  a network of  two $M/GI/\infty$ queues. Up to two transmissions attempts are allowed for all packets, i.e., $m_1=2$. Packets of sub-classes one and two are shown by red and blue colors, respectively. Observe that a packet will change  its sub-class after a decoding failure.  }
  \label{fig:queuing_system}
\end{figure*} 

\subsection{Infinite System Bandwidth }
Once again  consider a system with infinite system bandwidth so that there is no blocking of packets. In the multiple transmission model the BS has to wait for feedback from the intended receiver before re-transmitting a packet.    We  model this system with feedback using a network of two multi-class $M/GI/\infty$ queues, one modeling  BS transmissions and the other modeling the  packets awaiting feedback, which we refer to as the \emph{feedback} queue. 
%The packets arrive to the BS which is modeled as a multi-class $M/GI/\infty$ queue and they are transmitted to their intended receivers. After a packet gets decoded at its receiver, feedback is sent to the BS and  packet is re-transmitted if necessary. We model the packet decoding process using a \emph{feedback queue} which is also a multi-class $M/GI/\infty$ queue and appropriate packet routing.     
This is described below. 

{\em Base Station queue:} Similar to Sec.~\ref{sec:one_shot_transmission}, the BS is modeled as a multi-class $M/GI/\infty$ queue where each class corresponds to a set of users with the same SINR. However, unlike the one-shot transmission model, we further divide each class into various \emph{sub-classes} to keep track of the number of re-transmissions. In particular each class $c$ is further divided into $m_c$ sub-classes with the sub-classes indexed by various possible stages of packet (re)transmission. A class $c$ packet which is being transmitted for the $m^{\text{th}}$ time belongs to $m^{\text{th}}$ sub-class and it will require a bandwidth $h_{c,m}$  for a period of time $s_{c,m}$  to complete transmission. Further, because of our assumption of infinite bandwidth, the BS can transmit any number of packets from any of its classes concurrently, i.e., the number of servers in the queuing model is $\infty$. 

{\em Feedback queue:} We model the packet decoding and feedback sending processes as a multi-class  $M/GI/\infty $ queue which uses the same notion of a class and sub-class in the feedback queue as in the BS queue.    For  a class $c$ packet, the feedback associated with the decoding of a class $c$ packet is received at the BS after  $f_c$ seconds. Based on the success/failure of the decoding process  the BS  then decides to re-transmit it or not.  We abstract this process as follows. A  class $c$ packet after its $m^{\text{th}}$ transmission is routed from the BS queue to the feedback queue where it spends $f_c$ seconds. Note that the packet retains its class and sub-class indices in the feedback queue.  After $f_c$ seconds in the feedback queue it is then routed to the sub-class $m+1$ of  class $c$ with probability $p_{c,m}\brac{ \bvec{r}_c^{(m)}}$ (decoding failure) or leaves the system with probability $1-p_{c,m}\brac{ \bvec{r}_c^{(m)}}$ (successful decoding). If a class $c$ packet in $m^{\text{th}}$ sub-class is routed to the BS, then it changes its sub-class index to $m+1$, i.e., it is being transmitted for $(m+1)^{\text{th}}$ time.   This process repeats until the packet is successfully  decoded, or $m_c$ transmission attempts have been made, whichever happens first. Thus a  class $c$ packet always leaves the system after $m_c$  transmissions irrespective of the outcome of the  decoding process of the $m_c^{\text{th}}$ transmission. The queuing network  consisting of a BS and a single  class of URLLC traffic is illustrated in the Fig.~\ref{fig:queuing_system}.
  
 Observe that it is assumed that any number of URLLC packets can be processed in parallel in the feedback queue, and hence it can also be modeled as an $M/GI/\infty$ queue.  This is a reasonable assumption because the packet decoding process  across users are independent of each other and  done in parallel and we assume sufficient uplink bandwidth is provisioned for feedback from various users.

The queuing model described previously can be used to study various important properties of the multi-class system which are given below.  Let $\bvec{N}$ now be a random vector denoting the number of packets in different stages of re-transmissions of  all classes in the steady state, i.e., \\ $\bvec{N}:= \brac{N_{1,1}, N_{1,2} \ldots, N_{1,m_1},  \ldots, N_{C,1}, N_{C,2} \ldots, N_{C, m_C}}$.  Once again, from  classical queuing theory results (see~\cite{Harchol-Balter_book}), it follows that the steady state probability $\pi(\bvec{n})= P\brac{\bvec{N}=\bvec{n}}$ is given by:
\begin{equation}
\pi\brac{\bvec{n}} = \Pi_{c=1}^{C} \Pi_{m=1}^{m_c} \brac{ \frac{\rho_{c,m}^{n_{c,m}}}{n_{c,m} !}} \exp\brac{-\rho_{c,m}},
\end{equation}
where $\rho_{c,m}$ is the average system load of class $c$ packets in sub-class  $m$. 
%The  average packet transmission delay for class $c$ packets, denoted by $\tau_c$, is given by the following  expression
%\begin{equation}
%\tau_c = \frac{1}{\lambda_c} \sum_{m=1}^{m_c} \rho_{c,m} + f_c\brac{1 + \sum_{m=1}^{m_c} \Pi_{j=1}^m p_{c,j}\brac{\bvec{r}_c^{(j)}}},
%\end{equation}
%and  
The average bandwidth utilized is given by $\expect{\bvec{h}\bvec{N}^{T} }=\bvec{h}\boldsymbol{\rho}^{T}$.

%Therefore given $\bvec{M}$, $\bvec{h}$ and $\bvec{s}$, one can  characterize the system using the above result. 
%Next we shall describe about finding the optimal values of $\bvec{M}$, $\bvec{h}$ and $\bvec{s}$ such that the capacity of URLLC traffic is maximized. 
\subsection{Effect of Finite System Bandwidth }
Paralleling the one-shot transmission model studied earlier, a finite bandwidth system may suffer from congestion due to stochastic  variations in loads and may have to block an immediate packet transmission request (a new packet or a re-transmission). Hence, we must choose $W$ appropriately to meet the reliability requirements. A natural extension to the result in~\eqref{eq:multi_class_sqrt_staffing_rule_one_shot} for  the bandwidth requirement of the multi-class system is as  follows. Given  a target blocking probability of $\delta$, $W$ is chosen such that
\begin{equation}
\label{eq:multi_class_sqrt_staffing_rule}
    W \geq  {\eta}^{\text{mean}}\brac{\bvec{r},\bvec{m}} + c(\delta) \sqrt{{\eta}^{\text{variance}}\brac{\bvec{r},\bvec{m}, \bvec{h}, \bvec{s}}}, 
 \end{equation}  
where 
\begin{multline}
{\eta}^{\text{mean}}\brac{\bvec{r},\bvec{m}}:= \\ \frac{1}{\kappa} \sum_{c=1}^{C} \lambda_c \brac{  r_{c,1} +  \sum_{m=2}^{m_c} \brac{\Pi_{k=1}^{m-1} p_{c,k}\brac{\bvec{r}_c^{(k)}}}  r_{c,m}} 
\end{multline}
and
 \begin{multline}
{\eta}^{\text{variance}}\brac{\bvec{r},\bvec{m}, \bvec{h}, \bvec{s}}:= \\ \frac{1}{\kappa^2} \sum_{c=1}^{C} \lambda_c \brac{\frac{r_{c,1}^2}{s_{c,1}}  + \sum_{m=2}^{m_c}\brac{\Pi_{k=1}^{m-1} p_{c,k}\brac{\bvec{r}_c^{(k)}}} \frac{r_{c,m}^2}{s_{c,m}}}.
\end{multline}
Note that we have used the fact that $\kappa h_{c,m} s_{c,m}= r_{c,m}$ in writing the above equations.

The above characteristics follow by applying the square-root staffing rule to the multi-class system. The first term $\eta^{\text{mean}}\brac{\bvec{r},\bvec{m}}$  represents the mean bandwidth utilization. The term $\eta^{\text{variance}}\brac{\bvec{r},\bvec{m}, \bvec{h}, \bvec{s}}$ represents the variance of the bandwidth utilization.   Observe that while $\eta^{\text{mean}}\brac{\bvec{r},\bvec{m}}$ only depends on $\bvec{r}$ and $\bvec{m}$, each term in $\eta^{\text{variance}}$ is multiplied with  $1/s_{c,m}$ and thus is affected by the choice of $\bvec{s}$. 

%In addition to the blocking of  URLLC  packet transmissions at the BS  a URLLC transmission may suffer from decoding failure at the receiver after undergoing the maximum number of transmissions. For a class $c$ packet, this happens with probability $\Pi_{m=1}^{m_c}p_{c,m}\brac{\bvec{r}_c^{(m)}}$. Thus one would design the system such that $\Pi_{m=1}^{m_c}p_{c,m}\brac{\bvec{r}_c^{(m)}} \leq  \delta$.   We consider only transmission schemes which do not violate the deadline and if a packet is not successfully decoded when its deadline expires, then it is removed from the system. Hence, the probability of deadline violation is implicitly captured by the decoding failure probability after $m_c$ stages.

For $m_c=1$, we have shown in Thm.~\ref{thm:scaling_result} that  it is advantageous in terms of blocking probability to increase $s_{c,m}$ (or  decrease $h_{c,m}$) subject to the deadline constraint. This is not easily extendable to the case for $m_c >1$. However, this result gives us a key insight on the choice of $s_{c,m}$.  A natural extension of this insight to higher values of $m_c$ is to increase the transmission times of all stages such that the cumulative transmission time of $m_c$ stages and feedback delays add up to $d$, i.e.,
\begin{equation}
\label{eq:service_time_deadline}
 \sum_{m=1}^{m_c} s_{c,m} + m_c f_c =d. 
 \end{equation} 

% Recall that URLLC packets had to delivered within $d$ seconds with a probability of failure in delivering the packet to be less than $\delta$.
  Based on the previous discussion let us discuss the various steps one might follow to properly dimension this  multi-class system appropriately to support URLLC traffic.
\begin{enumerate}
\item Choose $\bvec{r}$ and $\bvec{m}$  such that probability of  decoding failure is less than or equal to $\delta$.
\item   Choose $\bvec{s}$ such that the condition~\eqref{eq:service_time_deadline} is satisfied. This also determines $\bvec{h}$ as $\bvec{r}$ is chosen in the first step and $\kappa h_{c,m} s_{c,m}=r_{c,m}$. 
\item To support an arrival rate vector $\boldsymbol{\lambda}$, determine the minimum necessary bandwidth via~\eqref{eq:multi_class_sqrt_staffing_rule}. 
  \end{enumerate}

  Although~\eqref{eq:multi_class_sqrt_staffing_rule} and~\eqref{eq:service_time_deadline} provide the basic insight into the effect of re-transmissions on the URLLC capacity, however there are still many possible solutions which satisfy~\eqref{eq:multi_class_sqrt_staffing_rule} and~\eqref{eq:service_time_deadline}.    One has to  find the optimal values for $\bvec{r}$, $\bvec{h}$, $\bvec{s}$, and $\bvec{m}$ to maximize the URLLC capacity. This is  discussed in the next section.

\section{URLLC Capacity Maximization/ Required Bandwidth Minimization}
\label{sec:urllc_capacity}
There are two ways to formulate the problem of optimizing  HARQ schemes to maximize URLLC capacity. One can characterize the set of URLLC arrival rates which can be supported  for a given system bandwidth $W$ subject to the QoS constraints. This  will define a \emph{multi-class URLLC capacity region}. Alternatively, one can formulate the problem in terms of minimizing the bandwidth required to support a given vector $\boldsymbol{\lambda}$ of URLLC arrival rates subject to the QoS constraints.  This second approach is somewhat simpler yet still allows one to study the most efficient system design  for the  HARQ schemes.    One can then study the structural properties of the solution obtained. We shall follow this second approach in the rest of this paper.  The associated  optimization problem is as follows:
 \begin{align}  
  \mathcal{OP}_2:\quad    \underset{\bvec{m}, \bvec{r},\bvec{h}, \bvec{s}}{\min: }& \,\,  {\eta}^{\text{mean}}\brac{\bvec{r},\bvec{m}} + c(\delta) \sqrt{{\eta}^{\text{variance}}\brac{\bvec{r},\bvec{m}, \bvec{h}, \bvec{s}}} \label{eq:objective_OP_2} \\
 \text{s.t.} \quad \kappa h_{c,m} s_{c,m}&=r_{c,m}, \,  \sum_{m=1}^{m_c} s_{c,m} + m_c f_c \leq d,   \\
 \text{and } h_c & \leq W, \, \Pi_{m=1}^{m_c} p_{c,m} \brac{\bvec{r}_c^{(m)}} \leq \delta, \quad \forall c.
 \end{align}
 
 The above problem is a non-convex, mixed integer programming problem, and in general is analytically intractable. To get some insights on this problem we will  consider a specific scheme, namely, \emph{repetition coding with homogeneous transmissions}. The performance under this scheme provides an upper bound on the minimum bandwidth required  under commonly used Chase combining, see~\cite{sesia_LTE}.   
%\subsection{Finite Block-Length  (FBL) Model}
%We require a analytical model which can quantify the resource requirements  for a given probability of  decoding failure. Since the packet sizes which are considered are very small, we shall use the capacity results in the finite blocklength regime developed in~\cite{Polyanskiy_2010}.  In an AWGN channel the number of information bits $L$ that can be transmitted  with a codeword decoding error  probability of $p$ in $r$ channel uses is given by 
%\begin{equation}
%\label{eq:block_length_most_exact}
%  L\approx r C(SINR)  -  Q^{-1} \brac{p} \sqrt{r V(SINR)} + 0.5 \log_{2}\brac{r} + o(1),
% \end{equation}  
%  where $C(SINR)= \log_2 \brac{1+SINR}$ is the AWGN channel capacity under infinite blocklength assumption $V(SINR)= \brac{\log_{2}(e)}^2\brac{1- \frac{1}{\brac{1+ SINR}^2}}$ and $Q(\cdot)$
%is the c.c.d.f. of the standard Gaussian distribution.   Using the above model we can now relate $h_{c,m}$, $s_{c,m}$, $r_{c,m}$ and $p_{c,m}$ for various HARQ mechanisms. We shall consider optimization of HARQ process for a simple repetition coding scheme in the next section. 
\subsection{Repetition Coding-- Homogeneous Transmissions}
 In repetition coding, the same codeword is transmitted repeatedly to the receiver until the packet is successfully decoded or the maximum number of re-transmissions has been reached. We shall also further  assume that the  transmissions are homogeneous. This is stated formally below. 
  \begin{assumption}
 \label{asm:static_bandwidth}
 ({\bf Homogeneous transmissions}) For all $c$ and  $m$, we have that $r_{c,m}=r_c$, $h_{c,m}=h_c $ and $s_{c,m}=s_c$.
 \end{assumption}
 We also make the following assumption on the packet decoding process at the receiver. 
 \begin{assumption}
 \label{asm:independent_decoding}
({\bf Independent decoding}) The receiver decodes each transmission independently of the previous transmissions, and hence, the probability of failure in any transmission attempt depends only on the codeword used in that stage. 
 \end{assumption}
 Under the above assumptions, the decoding failure probability is independent  across re-transmissions and driven by  $r_c$,  i.e.,  for any $c$ and $m$ we have that $p_{c,m}\brac{\bvec{r}_c^{(m)}}= p_{c}\brac{r_{c}}$.  Assuming independence between the decoding processes simplifies the analysis further. Also, due to the stringent latency requirements, complex HARQ schemes may not be practically feasible at the receiver. Utilizing homogeneous transmissions  reduces the overhead in control signals to indicate the allocation of bandwidth to users.
% \label{asm:constant_SINR}
% ({\bf Constant SINR across packet re-transmissions})  The SINR observed at any receiver is  a constant throughout a given packet transmission.   Let the SINR experienced by a class $c$ user be denoted by $SINR_c$. 
%\end{assumption} 
%The above assumption is reasonable because $d$ is typically less than $1$ msec and we can assume that the interference observed  changes at a slower time-scale. This assumption is also consistent with our assumption that users do not change class during an ongoing transmission in Assumption~\ref{asm:assumptions_on_multi_class_model}. 
%% Let $SINR_{c,m}$ be the SINR observed by a class $c$ user after receiving $m^{\text{th}}$ transmission of a packet. For chase combining,  under the constant SINR assumption, using the result in~\cite{} we have that 
%%\begin{equation}
%% SINR_{c,m}= mSINR_{c,1}.
% \end{equation} 
% Therefore, the SINR increases linearly with the number of transmissions in chase combining.

 Unfortunately, under finite block length model and repetition coding, $\mathcal{OP}_2$ is still analytically intractable in a  multi-class system. Therefore, we shall consider two  regimes,   \emph{the variance }  and  \emph{the mean utilization dominated regimes} where the solutions simplify considerably. They are formally described next.
 \begin{definition}
  ~ \hspace{-0.1in}
 \begin{enumerate}
 \item \emph{Variance dominated regime:} In this regime, the objective function of $\mathcal{OP}_2$ includes only  the variance of the bandwidth utilization ( $\eta^{\text{variance}}$). 
 \item \emph{Mean utilization dominated regime:} In this regime, the objective function $\mathcal{OP}_2$ includes only the mean of the bandwidth utilization ($\eta^{\text{mean}}$). 
 \end{enumerate}
 \end{definition}

  Note that at low loads when $\lambda_c$'s are  small,  in~\eqref{eq:objective_OP_2} the term corresponding to the overall  variance is dominant, therefore, at low loads we shall minimize the variance of the total bandwidth usage.  At high loads, the variance of bandwidth usage (second term) is smaller than the mean (first term). Hence, we shall focus on minimizing the mean utilization at high loads. We shall also use the finite blocklength model  discussed in Sec.~\ref{subsec:fbl} to relate $p_c(r_c)$ and $r_c. $  Under these simplifications, one can de-couple $\mathcal{OP}_2$ for each class and  optimize the HARQ schemes separately for each class.   The main result in the variance dominated regime is given below.
\begin{proposition}
\label{res:variance_driven_regime}
For the multiple transmissions model in Sec.~\ref{sec:retransmissions},  under Assumptions~\ref{asm:assumptions_on_multi_class_model},~\ref{asm:static_bandwidth}, and~\ref{asm:independent_decoding}, and in the variance dominated regime, the optimization problem $\mathcal{OP}_2$  decomposes across classes. The optimization problem for class $c$ is as follows:
\begin{align}
\underset{m_c, r_c, h_c, s_c}{\min:} & \sum_{m=0}^{m_c}  \brac{\frac{r_{c}^2}{s_{c}}}  {\brac{p_c(r_c)}^{m}}  \\
\text{s.t.} \quad   \kappa s_c h_c&=r_c, \, \, h_c  \leq W, \, \, m_c\brac{ s_{c} +  f_c} = d, \\
  &  \brac{p_c(r_c)}^{m_c}\leq \delta. 
\end{align} 
Furthermore, under the finite block length model~\eqref{eq:block_length_approximation} relating $p_c(r_c)$ and $r_c$, for $L \leq 2000$ bits, $d \leq 2$ msec., $\delta \in \sbrac{10^{-3}, 10^{-7}}$, $SINR_c \in \sbrac{0, 20} dB$, $f_c \geq 0.1$ msec. the optimal solution has the following structure:
\begin{enumerate}
\item One shot transmission is optimal, i.e., $m_c^*=1$ .
\item The optimal values of transmission time and bandwidth, denoted by $s_c^*$ and $h_c^*$, respectively, satisfy
\begin{equation}
s_c^*=d-f_c \text{ and }  h_c^* = \frac{r_c^*}{d-f_c},
\end{equation}
where $r_c^*$ is the smallest $r$ such that $p_c(r)\leq\delta$. 
\end{enumerate}
\end{proposition}
%A proof of this result is given in Appendix~\ref{pf:variance_driven_regime}.

% In the variance driven regime, we have to solve the following optimization problem. 
%  \begin{align}
% \mathcal{OP}_3: \quad \underset{M_c, h_c, s_c}{\min:} & \sum_{c=1}^{C} \lambda_{c} \sum_{m=0}^{M_c}  h_{c}^2  s_{c}  {p_c^{m}}  \\
% \text{s.t.} \quad h_{c} s_{c}&=r_{c} \quad \forall c, \label{eq:resource_conservation} \\
% M_c s_{c} + M_c f_c &\leq d, \quad \forall c, \label{eq:total_time}\\
% h_c &\leq W, \quad \forall c, \\
% p_c^{M_c}&\leq \delta. 
% \end{align}
% Observe that the above problem can be decomposed into separate sub-problems for each class which can be solved independently. The main result in this regime is given below.
% \begin{result}
% \label{res:variance_driven_regime}
% In the variance driven regime,  under Assumptions~\ref{asm:assumptions_on_multi_class_model},~\ref{asm:constant_SINR}, and~\ref{asm:static_bandwidth},  the optimal solution to  $\mathcal{OP}_3$ for a range of packet sizes and SINRs is a  one-shot transmission  with the following parameters:
% \begin{align}
% s_c&=d-f_c \quad \forall c, \\
% h_c &= \frac{r_c}{d-f_c}. 
% \end{align}
% \end{result} 
% An explanation with a numerical proof of this result is given in Appendix~\ref{}.   

The equivalent result in the mean utilization dominated regime is given below. 
\begin{proposition}
\label{res:mean_util_driven_regime}
For the  multiple transmission model in Sec.~\ref{sec:retransmissions}, under Assumptions~\ref{asm:assumptions_on_multi_class_model},~\ref{asm:static_bandwidth}, and~\ref{asm:independent_decoding}, and in the mean utilization dominated regime, the optimization problem $\mathcal{OP}_2$   decomposes across classes.  The optimization for class $c$ is given by:
\begin{align}
\underset{m_c, r_c, h_c, s_c}{\min:} & \sum_{m=0}^{m_c}  r_c  {\brac{p_c(r_c)}^{m}}  \\
\text{s.t.} \quad  \kappa s_c h_c&=r_c, \,\, h_c \leq W, \,\, m_c\brac{ s_{c} +  f_c} = d, \\
   &\brac{p_c(r_c)}^{m_c} \leq \delta. 
\end{align} 
Furthermore, under the finite block length model~\eqref{eq:block_length_approximation} relating $r_c$ and $p_c(r_c)$, for $L \leq 2000$ bits, $d \leq 2$ msec., $\delta \in \sbrac{10^{-3}, 10^{-7}}$, $SINR_c \in \sbrac{0, 20} dB$, $f_c \in \sbrac{0.1, 0.25}$ msec. the optimal solution has the following structure:
\begin{enumerate}
\item The optimal value  $m_c^*$ is strictly more than one.
\item The optimal value  $m_c^*$ is a non-increasing function of $SINR_c$.
\end{enumerate}
\end{proposition}
%A proof of this result is given in Appendix~\ref{pf:mean_util_driven_regime}. 
The decomposed optimization problems for  class $c$ in both the regimes are obtained from the  definitions of $\eta^{\text{mean}}\brac{\bvec{r}, \bvec{m}}$ and $\eta^{\text{variance}}\brac{\bvec{r}, \bvec{m}, \bvec{h}, \bvec{s}}$. The results on $m_c^*$ are obtained by direct substitution. Some remarks regarding the above two propositions are in order. 

\begin{figure}
\centering
\includegraphics[scale=0.65]{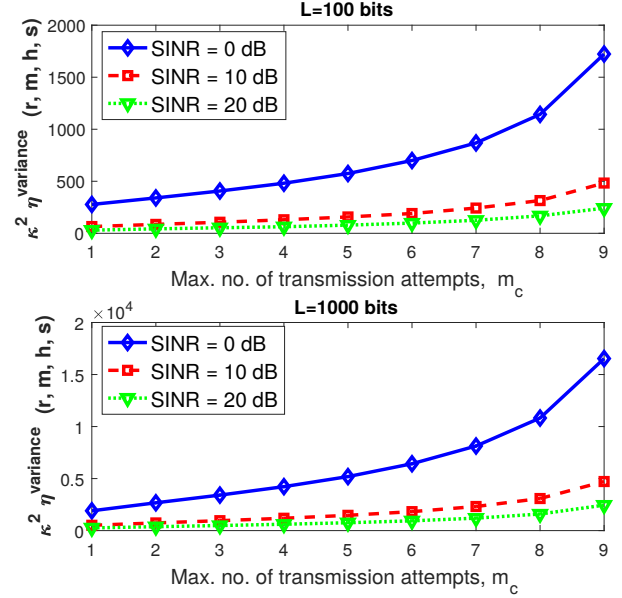}
\caption{Variance of bandwidth utilization (scaled) as a function of $m_c$ for various values of $SINR_c$ and $L$ with $\lambda_c=1 $ packet/msec., $\delta=10^{-6}$, $d=1$ msec.,  and $f_c=0.125$ msec. }
\label{fig:static_bandwidth_allocation_variance}
\end{figure}

\begin{figure}
\centering
\includegraphics[scale=0.65]{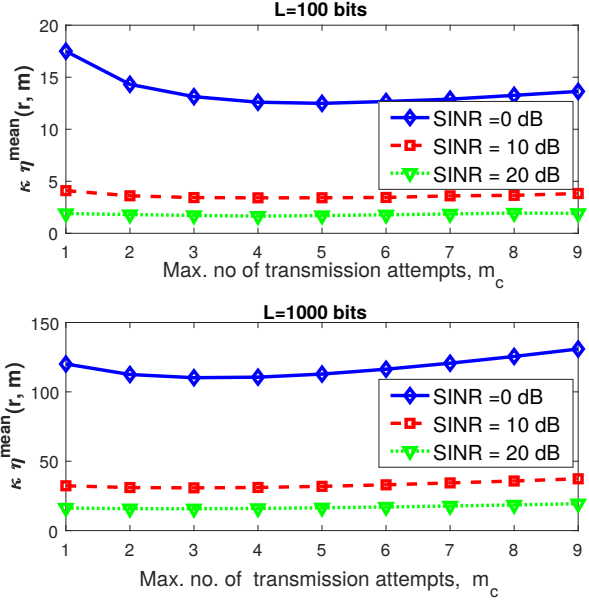}
\caption{Mean of bandwidth utilization (scaled) as a function of $m_c$ for various values of $SINR_c$  and $L$ with $\lambda_c=100 $ packet/msec., $\delta=10^{-6}$, $d=1$ msec.,  and $f_c=0.125$ msec.}
\label{fig:static_bandwidth_allocation_mean}
\end{figure}

Comparing the objective functions for the mean and variance dominated regimes, note that each term in the variance dominated regime is multiplied with an extra term $\frac{r_c}{ s_c}$. Since $\frac{r_c}{ s_c}=\frac{r_cm_c}{\brac{d-m_cf_c}}$, the objective function in the variance dominated regime increases sharply with  $m_c$. Therefore, the optimal value of $m_c$ is lower in the variance dominated regime than in the mean dominated regime. In the mean dominated regime, as one decreases $SINR_c$, the resources required per transmission ($r_c$) to meet a given reliability requirement increase sharply.  Hence, it is advantageous at lower SINRs to increase $m_c$ and to choose a lower reliability target per transmission. We have plotted the variance and mean of bandwidth utilization for  various packet sizes and SINR values in Figures~\ref{fig:static_bandwidth_allocation_variance} and~\ref{fig:static_bandwidth_allocation_mean}, respectively. 
 
\section{Conclusions}
\label{sec:conclusions}
In this paper we explored possible designs of 5G wireless systems supporting URLLC traffic.
We develop a simple model for URLLC packet transmissions which captures the 
essential properties of such a system when pre-emptive/immediate URLLC scheduling and finite
block-length transmissions are used.  Based on this model we derive
scaling results for the URLLC capacity (admissible load subject to QoS constraints) with 
respect to various system parameters such as the link SINR, system bandwidth, and the 
packet latency and reliability requirements. Several key findings arise which are of practical interest. 
First, URLLC capacity is enhanced by extending URLLC transmissions in time as much as possible (subject
to latency constraints) while using the least amount of bandwidth (to meet reliability requirements).
Next we consider results associated with  optimizing   HARQ schemes.  
In the variance dominated regime (typically low loads), one-shot transmissions satisfying the above mentioned requirements minimize the necessary bandwidth required to support URLLC traffic.  In the mean utilization  dominated regime (typically high loads), optimal HARQ schemes minimizing the necessary bandwidth leverage multiple re-transmissions and the maximum number of transmissions required is a non-increasing function of SINR. 
%\begin{enumerate}
%\item Extending transmissions  in time as much as possible while using the least amount of bandwidth to meet the latency constraints increases the URLLC capacity.  
%\item At low URLLC loads, a one-shot transmission spread out as much as possible in time and meeting the required reliability requirement  maximizes the spectral efficiency.
%\item At high URLLC loads, the  FEC and HARQ scheme  maximizing the spectral efficiency permits multiple transmissions to meet the desired reliability requirement. If we use Chase combining, the optimal  maximum number of transmissions permitted is a non-increasing function SINR. Whereas if we use Incremental Redundancy  scheme, the optimal maximum number of transmissions permitted is two and is independent of SINR. Further it is designed such that the probability of decoding failure after first stage is approximately $10^{-2}$.          
%\end{enumerate}

% Using this queuing network model we have studied the effect of various parameters like packet size, SINR, FEC and HARQ scheme,  deadline constraint, and  reliability factor  on the URLLC capacity. We have also optimized practically relevant FEC and HARQ schemes to maximize the URLLC capacity.    
\appendices

\subsection{Proof of Theorem~\ref{thm:multi_class_block_prob}}
\label{pf:multi_class_block_prob}
 Without loss of generality, let us consider $\blockprobclass{1}{\bvec{h}}{\bvec{s}}{\boldsymbol{\lambda}}{W}$.  Using the standard results from queuing theory (see~\cite{Harchol-Balter_book}), we have that:
\begin{equation}
\pi(\bvec{n})= G \,  {\Pi_{c=1}^{C} \brac{\frac{\rho_c^{n_c}}{n_c!}}},
\end{equation}
where $G^{-1} = \sum_{\tilde{\bvec{n}} \in \mathcal{S}} \Pi_{c=1}^{C} \brac{\frac{\rho_c^{\tilde{n}_c}}{\tilde{n}_c!}}  $ and $\mathcal{S}=\cbrac{\bvec{n} \mid \bvec{h}\bvec{n}^{T} \leq W}$. Here $\mathcal{S}$ is the set of all user configurations such that the total bandwidth constraint is not violated. Similarly define $\mathcal{S}'=\cbrac{\bvec{n} \mid \bvec{h}'\bvec{n}^{T} \leq W}$. From the definition of $\bvec{h}$ and $\bvec{h}'$, we have that 
\begin{equation}
\label{eq:relating_two_systems}
\bvec{n} \in \mathcal{S} \Leftrightarrow \sbrac{n_1, n_2, \ldots, qn_i, \ldots, n_C} \in \mathcal{S}'. 
\end{equation}

Define $\mathcal{S}_1:=\cbrac{\bvec{n} \mid \bvec{n} \in \mathcal{S} \text{ and } \bvec{n} + \bvec{e}_1  \notin \mathcal{S}}$, where $\bvec{e}_1$ is the unit vector with only the first coordinate as the non-zero element. $\mathcal{S}_1$ is the set of states in which class $1$ users experience blocking. Similarly define $\mathcal{S}_1'$ for the case with bandwidths $\bvec{h}'$ and $\bvec{s}'$. Observe that due to~\eqref{eq:relating_two_systems} we have   $\mathcal{S} \subseteq \mathcal{S}'$. Furthermore, if $\bvec{n} \in \mathcal{S}_1$, then for $\tilde{n}_i \in \cbrac{q n_i -\ceil{\frac{h_1 q}{h_i}} +1, q n_i -\ceil{\frac{h_1 q}{h_i}} +2, \ldots, q n_i}$ we have that  $\bvec{n}':= \sbrac{n_1, n_2, \ldots, \tilde{n}_i, \ldots, n_C} \in \mathcal{S}_1'$.  Using PASTA property (see~\cite{Harchol-Balter_book}), the blocking probability experienced by a typical arrival to class $1$ is given by 
\begin{equation}
\label{eq:block_prob_multi-class_baseline}
 \blockprobclass{1}{\bvec{h}}{\bvec{s}}{\boldsymbol{\lambda}}{W} = \frac{\sum_{\bvec{n} \in \mathcal{S}_1} {\Pi_{c=1}^{C} \brac{\frac{\rho_c^{n_c}}{n_c!}}}  }{\sum_{\bvec{n} \in \mathcal{S}} {\Pi_{c=1}^{C} \brac{\frac{\rho_c^{n_c}}{n_c!}}}}.  
   \end{equation}   
Similarly, the blocking probability experienced under $\bvec{h}'$ and $\bvec{s}'$ is given by
\begin{equation}
\label{eq:block_prob_multi-class_scaled_system}
 \blockprobclass{1}{\bvec{h}'}{\bvec{s}'}{\boldsymbol{\lambda}}{W} = \frac{\sum_{\bvec{n} \in \mathcal{S}'_1} q^{n_i} {\Pi_{c=1}^{C} \brac{\frac{\rho_c^{n_c}}{n_c!}}}   }{\sum_{\bvec{n} \in \mathcal{S}'} q^{n_i} {\Pi_{c=1}^{C} \brac{\frac{\rho_c^{n_c}}{n_c!}} }}.  
\end{equation}
We will show that $ \blockprobclass{1}{\bvec{h}'}{\bvec{s}'}{\boldsymbol{\lambda}}{W} \leq \blockprobclass{1}{\bvec{h}}{\bvec{s}}{\boldsymbol{\lambda}}{W}$. We can re-write~\eqref{eq:block_prob_multi-class_baseline} as follows:
\begin{equation}
\label{eq:block_prob_multi-class_baseline_rearranged}
\blockprobclass{1}{\bvec{h}}{\bvec{s}}{\boldsymbol{\lambda}}{W} = \frac{1}{ 1+ \frac{\sum_{\bvec{n} \in \mathcal{S} \diagdown \mathcal{S}_1} {\Pi_{c=1}^{C} \brac{\frac{\rho_c^{n_c}}{n_c!}}}}{\sum_{\bvec{n} \in \mathcal{S}_1} \Pi_{c=1}^{C} \brac{\frac{\rho_c^{n_c}}{n_c!}}}}.
\end{equation}

Next we will re-write~\eqref{eq:block_prob_multi-class_scaled_system} as follows:
\begin{multline}
\label{eq:block_prob_multi-class_scaled_system_rearranged}
\blockprobclass{1}{\bvec{h}'}{\bvec{s}'}{\boldsymbol{\lambda}}{W}  = \\ \frac{1}{ 1+ \frac{\sum_{\bvec{n} \in \mathcal{S} \diagdown \mathcal{S}_1} q^{n_i} {\Pi_{c=1}^{C} \brac{\frac{\rho_c^{n_c}}{n_c!}} }}{\sum_{\bvec{n} \in \mathcal{S}'_1} q^{n_i} \Pi_{c=1}^{C} \brac{\frac{\rho_c^{n_c}}{n_c!}} } +   \frac{\sum_{\bvec{n} \in \mathcal{S'} \diagdown {\mathcal{S} \diagdown \mathcal{S}_1}} q^{n_i} {\Pi_{c=1}^{C} \brac{\frac{\rho_c^{n_c}}{n_c!}} }}{\sum_{\bvec{n} \in \mathcal{S}'_1} q^{n_i} \Pi_{c=1}^{C} \brac{\frac{\rho_c^{n_c}}{n_c!}} }}
 \end{multline} 
 To compare $ \blockprobclass{1}{\bvec{h}'}{\bvec{s}'}{\boldsymbol{\lambda}}{W}$ and $\blockprobclass{1}{\bvec{h}}{\bvec{s}}{\boldsymbol{\lambda}}{W}$, let us compare the denominators of~\eqref{eq:block_prob_multi-class_baseline_rearranged} and~\eqref{eq:block_prob_multi-class_scaled_system_rearranged}. We will show the following:
 \begin{equation}
 \label{eq:main_inequality}
\frac{\sum_{\bvec{n} \in \mathcal{S} \diagdown \mathcal{S}_1} q^{n_i} {\Pi_{c=1}^{C} \brac{\frac{\rho_c^{n_c}}{n_c!}} }}{\sum_{\bvec{n} \in \mathcal{S}'_1} q^{n_i} \Pi_{c=1}^{C} \brac{\frac{\rho_c^{n_c}}{n_c!}} }  \geq \frac{\sum_{\bvec{n} \in \mathcal{S} \diagdown \mathcal{S}_1} {\Pi_{c=1}^{C} \brac{\frac{\rho_c^{n_c}}{n_c!}}}}{\sum_{\bvec{n} \in \mathcal{S}_1} \Pi_{c=1}^{C} \brac{\frac{\rho_c^{n_c}}{n_c!}}}.
 \end{equation}
If the above equation holds, then from~\eqref{eq:block_prob_multi-class_baseline_rearranged} and~\eqref{eq:block_prob_multi-class_scaled_system_rearranged}, it can be easily shown that $ \blockprobclass{1}{\bvec{h}'}{\bvec{s}'}{\boldsymbol{\lambda}}{W} \leq \blockprobclass{1}{\bvec{h}}{\bvec{s}}{\boldsymbol{\lambda}}{W}$.  Note that in the above expression the numerator of the L.H.S. is greater than the numerator of the R.H.S. Next we have to compare the denominators. Due to~\eqref{eq:relating_two_systems},  we can re-write the denominator of the L.H.S. as follows:
 \begin{multline}
\sum_{\bvec{n} \in \mathcal{S}'_1}q^{n_i} \Pi_{c=1}^{C} \brac{\frac{\rho_c^{n_c}}{n_c!}}  \\  = \sum_{\bvec{n} \in \mathcal{S}_1} \sum_{i=1}^{\frac{q h_1}{h_i}} \frac{\brac{q\rho_i}^{\brac{qn_i - \ceil{\frac{q h_1}{h_i}} + i}}}{\brac{qn_i - \ceil{\frac{q h_1}{h_i}} + i} !}\Pi_{c: c\neq i} \brac{\frac{\rho_c^{n_c}}{n_c!}} 
  \end{multline} 
  It can be shown that in wide-band systems with $W$  large enough and $\rho_i <  1$ the following holds: 
  \begin{equation}
\sum_{\bvec{n} \in \mathcal{S}_1} \sum_{i=1}^{\frac{q h_1}{h_i}} \frac{\brac{q\rho_i}^{\brac{qn_i - \ceil{\frac{q h_1}{h_i}} + i}}}{\brac{qn_i - \ceil{\frac{q h_1}{h_i}} + i} !}\Pi_{c: c\neq i} \brac{\frac{\rho_c^{n_c}}{n_c!}}  \leq  \sum_{\bvec{n} \in \mathcal{S}_1} \Pi_{c=1}^{C} \brac{\frac{\rho_c^{n_c}}{n_c!}}.
  \end{equation}
  Therefore, the denominator of the L.H.S. of~\eqref{eq:main_inequality} is less than the denominator of its R.H.S. We have proved the inequality~\eqref{eq:main_inequality}, and hence, $ \blockprobclass{1}{\bvec{h}'}{\bvec{s}'}{\boldsymbol{\lambda}}{W} \leq \blockprobclass{1}{\bvec{h}}{\bvec{s}}{\boldsymbol{\lambda}}{W}$.  
%\subsection{Proof of Result~\ref{rs:result_on_queuing_system}}
%\label{pf:result_on_queuing_system} 
%In this analysis we shall assume that the packets are never blocked. Since $\delta$ is typically very small ($10^{-6}$) this is a good approximation. A packet is transmitted for the $m+1^{\text{th}}$ time if the transmission fails in the first $m$ attempts. For a class $c$ packet, this happens with probability $\Pi_{m=0}^{m'} p_{c,m'}$. Therefore the overall rate at which a class $c$ packet in $m+1^{\text{th}}$ transmission stage enters the knapsack is $\lambda_c\Pi_{m'=0}^{m} p_{c,m'}$ and  it bring a load of $\rho_{c,m}$ to the system.   Once we have the result follows from the standard results on Jackson networks, see Chapter~~\cite{kleinrockI75_book} for more details.    
\subsection{Approximate Expression for Blocklength}
\label{pf:approx_expression_for_r}
If we ignore the terms $0.5\log_2 (r)$  and $o(1)$  in~\eqref{eq:block_length_most_exact}, we have the following approximate expression relating blocklength $r$, the number of information bits $L$ and the probability of  decoding failure $p$.
\begin{equation}
  L\approx r C(SINR_c)  -  Q^{-1} \brac{p} \sqrt{r V(SINR_c)}.
 \end{equation}
 If we substitute $\sqrt{r}=x$, then the above equation is a quadratic equation in $x$. Solving it we get the approximate expression for $r$ in~\eqref{eq:block_length_approximation}.  
 %Plots showing the accuracy of this approximation are given in Chapter~\ref{},~\cite{}. 

% In Fig.~\ref{fig:approximation_of_block_length} we have compared the values of $r$ obtained from expression~\eqref{eq:block_length_most_exact} and  with the our approximation~\eqref{eq:block_length_approximation} for different packet sizes, SINRs and probability of decoding failure. Both the expressions give almost similar values of  blocklengths.
% 
% \begin{figure}
% \begin{subfigure}{0.5\textwidth}
% \centering
%  \includegraphics[width=\linewidth]{block_approx_0dB}
%  \caption{}
%  \label{fig:block_approx_0dB}
% \end{subfigure}
% %
%  \begin{subfigure}{0.5\textwidth}
% \centering
%  \includegraphics[width=\linewidth]{block_approx_10dB}
%  \caption{}
%   \label{fig:block_approx_10dB}
% \end{subfigure}
% \begin{subfigure}{0.5\textwidth}
% \centering
%  \includegraphics[width=\linewidth]{block_approx_20dB}
%  \caption{}
%   \label{fig:block_approx_20dB}
% \end{subfigure}
% \caption{Comparison $r$ obtained using our approximation~\eqref{eq:block_length_approximation} with respect to the expression for blocklength derived in~\cite{Polyanskiy_2010} and re-stated in~\eqref{eq:block_length_most_exact}. } 
% \label{fig:approximation_of_block_length}
% \end{figure}
 
\subsection{Proof of Theorem~\ref{thm:scaling_result}}
\label{pf:scaling_result} 
From~\eqref{eq:multi_class_sqrt_staffing_rule}, we have the following relation between $\lambda_c^*$ and $W$
 \begin{equation}
 \label{eq:bandwidth_1}
\kappa W=\lambda_c^* r_{c} + c(\delta) r_{c}\sqrt{\frac{\lambda_c^*}{d}  },
\end{equation} 
where $r_{c}$, $L$, $SINR_c$,  and $\delta$ are related according to~\eqref{eq:block_length_approximation}, and the expression for $r_c$ is re-stated below:
\begin{multline}
r_{c} = \frac{L}{C(SINR_c)}  + \frac{\brac{Q^{-1}\brac{\delta}}^2V(SINR_c)}{2\brac{C(SINR_c)}^2}  \\ + \frac{\brac{Q^{-1}\brac{\delta}}^2V(SINR_c)}{2\brac{C(SINR_c)}^2} \sqrt{1 + \frac{4LC(SINR_c)}{V(SINR_c) \brac{Q^{-1}\brac{\delta}}^2 }}.
\end{multline}
% Substituting $h_c=\frac{r_c}{\kappa d}$ in~\eqref{eq:bandwidth_1}, we get 
%\begin{equation}
%W=\frac{1}{}\lambda_c^* r_{c} + c(\delta) r_{c} \sqrt{\frac{\lambda_c^*}{d}}. 
%\end{equation}
Solving for $\lambda_c^*$ in~\eqref{eq:bandwidth_1}, we have that 
\begin{equation}
\label{eq:expression_for_max_lambda}
\lambda_c^* = \frac{\kappa W}{r_c} + \frac{c(\delta)^2 }{d} \brac{1-\sqrt{1+\frac{4 \kappa W d}{c(\delta)^2 r_c}}}. 
\end{equation}
Scaling with respect to $W$ directly follows from~\eqref{eq:expression_for_max_lambda}. 

To understand the scaling with respect to $SINR_c$, we have to first study the scaling of $r_c$ with respect to $SINR_c$. For large $SINR_c$, we have that 
\begin{align}
C(SINR_c) &\sim \Theta\brac{\log_2 \brac{SINR_c}}, \\
V(SINR_c) &\sim \Theta\brac{1}. 
\end{align}
Therefore, $r_c \sim \Theta \brac{\frac{1}{\log_2 \brac{SINR_c}}}$. Using~\eqref{eq:expression_for_max_lambda}, we get that $\lambda_c^* \sim \Theta \brac{\log_2 \brac{SINR_c}  - \sqrt{ \log_2 \brac{SINR_c}}}  $. Similarly, using~\eqref{eq:expression_for_max_lambda}, we get the scaling with respect to $d$ as $\lambda_c^* \sim \Theta \brac{1- \frac{1}{\sqrt{d}}}$. If we use the square-root staffing rule with the normal approximation (see~\cite{Harchol-Balter_book}), we have that $c(\delta) = Q^{-1} \brac{\delta}\sim \Theta \brac{\sqrt{-\log\brac{\delta}}}$.  As we increase $\delta$, we $c\brac{\delta} \rightarrow 0$.   Using $Q^{-1} \brac{\delta}\sim \Theta \brac{\sqrt{-\log\brac{\delta}}}$ we have that $r_c \sim \Theta\brac{-\log \brac{\delta}}$. Therefore, from~\eqref{eq:expression_for_max_lambda} we get that $\lambda_c^* \sim  \Theta\brac{\frac{1}{-\log \brac{\delta}}}$.

\ifCLASSOPTIONcaptionsoff
  \newpage
\fi



%\bibliographystyle{IEEEtran}
\bibliographystyle{IEEEtran}
\bibliography{bibJournalList,arjun,ss-3}

\begin{thebibliography}{10}
\providecommand{\url}[1]{#1}
\csname url@samestyle\endcsname
\providecommand{\newblock}{\relax}
\providecommand{\bibinfo}[2]{#2}
\providecommand{\BIBentrySTDinterwordspacing}{\spaceskip=0pt\relax}
\providecommand{\BIBentryALTinterwordstretchfactor}{4}
\providecommand{\BIBentryALTinterwordspacing}{\spaceskip=\fontdimen2\font plus
\BIBentryALTinterwordstretchfactor\fontdimen3\font minus
  \fontdimen4\font\relax}
\providecommand{\BIBforeignlanguage}[2]{{%
\expandafter\ifx\csname l@#1\endcsname\relax
\typeout{** WARNING: IEEEtran.bst: No hyphenation pattern has been}%
\typeout{** loaded for the language `#1'. Using the pattern for}%
\typeout{** the default language instead.}%
\else
\language=\csname l@#1\endcsname
\fi
#2}}
\providecommand{\BIBdecl}{\relax}
\BIBdecl

\bibitem{hwwtahaa16}
B.~Holfeld, D.~Wieruch, T.~Wirth, L.~Thiele, S.~A. Ashraf, J.~Huschke,
  I.~Aktas, and J.~Ansari, ``Wireless communication for factory automation: an
  opportunity for {LTE} and {{5G}} systems,'' \emph{IEEE Communications
  Magazine}, vol.~54, no.~6, pp. 36--43, June 2016.

\bibitem{ywjbas15}
O.~N.~C. Yilmaz, Y.~P.~E. Wang, N.~A. Johansson, N.~Brahmi, S.~A. Ashraf, and
  J.~Sachs, ``Analysis of ultra-reliable and low-latency {5G} communication for
  a factory automation use case,'' in \emph{2015 IEEE International Conference
  on Communication Workshop (ICCW)}, June 2015, pp. 1190--1195.

\bibitem{gla17}
M.~Gidlund, T.~Lennvall, and J.~Akerberg, ``Will {5G} become yet another
  wireless technology for industrial automation?'' in \emph{IEEE Int. Conf. on
  Industrial Technology (ICIT)}, March 2017, pp. 1319--1324.

\bibitem{pbfms16}
K.~I. Pedersen, G.~Berardinelli, F.~Frederiksen, P.~Mogensen, and A.~Szufarska,
  ``A flexible {5G} frame structure design for frequency-division duplex
  cases,'' \emph{IEEE Communications Magazine}, vol.~54, no.~3, pp. 53--59,
  March 2016.

\bibitem{3gpp_ran1_87}
3GPP TSG RAN WG1 Meeting 87, November 2016.

\bibitem{3gpp_ran1_88}
Chairman's notes 3GPP: 3GPP TSG RAN WG1 Meeting 88bis, Available at
  \url{http://www.3gpp.org/ftp/TSG_RAN/WG1_RL1/TSGR1_88b/Report/}, April 2017.

\bibitem{Bennis_2018}
M.~{Bennis}, M.~{Debbah}, and H.~V. {Poor}, ``{Ultra-Reliable and Low-Latency
  Wireless Communication: Tail, Risk and Scale},'' \emph{ArXiv e-prints}, Jan.
  2018.

\bibitem{Popovski_2017}
\BIBentryALTinterwordspacing
P.~Popovski, J.~J. Nielsen, C.~Stefanovic, E.~de~Carvalho, E.~G. Str{\"{o}}m,
  K.~F. Trillingsgaard, A.~Bana, D.~Kim, R.~Kotaba, J.~Park, and R.~B.
  S{\o}rensen, ``Ultra-reliable low-latency communication {(URLLC):} principles
  and building blocks,'' \emph{CoRR}, vol. abs/1708.07862, 2017. [Online].
  Available: \url{http://arxiv.org/abs/1708.07862}
\BIBentrySTDinterwordspacing

\bibitem{Ji_2017}
\BIBentryALTinterwordspacing
H.~Ji, S.~Park, J.~Yeo, Y.~Kim, J.~Lee, and B.~Shim, ``Introduction to ultra
  reliable and low latency communications in 5g,'' \emph{CoRR}, vol.
  abs/1704.05565, 2017. [Online]. Available:
  \url{http://arxiv.org/abs/1704.05565}
\BIBentrySTDinterwordspacing

\bibitem{Ashraf_2017}
S.~Ashraf, Y.~P.~E. Wang, S.~Eldessoki, B.~Holfeld, D.~Parruca, M.~Serror, and
  J.~Gross, ``From radio design to system evaluations for ultra-reliable and
  low-latency communication,'' in \emph{Proc. European Wireless Conference},
  May 2017, pp. 1--8.

\bibitem{ljcjs17}
C.-P. Li, J.~Jiang, W.~Chen, T.~Ji, and J.~Smee, ``{5G} ultra-reliable and
  low-latency systems design,'' in \emph{2017 European Conference on Networks
  and Communications (EuCNC)}, June 2017, pp. 1--5.

\bibitem{dkp16}
G.~Durisi, T.~Koch, and P.~Popovski, ``Toward massive, ultrareliable, and
  low-latency wireless communication with short packets,'' \emph{Proceedings of
  the IEEE}, vol. 104, no.~9, pp. 1711--1726, Sept 2016.

\bibitem{Anand_2017}
A.~{Anand}, G.~{de Veciana}, and S.~{Shakkottai}, ``{Joint Scheduling of URLLC
  and eMBB Traffic in 5G Wireless Networks},'' \emph{ArXiv e-prints}, Dec.
  2017.

\bibitem{You_2018}
L.~{You}, Q.~{Liao}, N.~{Pappas}, and D.~{Yuan}, ``{Resource Optimization with
  Flexible Numerology and Frame Structure for Heterogeneous Services},''
  \emph{ArXiv e-prints}, Jan. 2018.

\bibitem{Shariatmadari_2015}
H.~Shariatmadari, S.~Iraji, and R.~Jantti, ``Analysis of transmission methods
  for ultra-reliable communications,'' in \emph{Proc.\ {PIMRC}}, Aug. 2015, pp.
  2303--2308.

\bibitem{Shariatmadari_2016}
H.~Shariatmadari, S.~Iraji, Z.~Li, M.~A. Uusitalo, and R.~Jantti, ``Optimized
  transmission and resource allocation strategies for ultra-reliable
  communications,'' in \emph{Proc.\ {PIMRC}}, Sep. 2016, pp. 1--6.

\bibitem{Malak_2017}
D.~{Malak}, H.~{Huang}, and J.~G. {Andrews}, ``{Throughput Maximization for
  Delay-Sensitive Random Access Communication},'' \emph{ArXiv e-prints}, 2017.

\bibitem{dkopy16}
G.~Durisi, T.~Koch, J.~Ostman, Y.~Polyanskiy, and W.~Yang, ``Short-packet
  communications over multiple-antenna rayleigh-fading channels,'' \emph{IEEE
  Trans. on Comm.}, vol.~64, no.~2, pp. 618--629, Feb 2016.

\bibitem{sltum16}
B.~Singh, Z.~Li, O.~Tirkkonen, M.~A. Uusitalo, and P.~Mogensen,
  ``Ultra-reliable communication in a factory environment for {5G} wireless
  networks: Link level and deployment study,'' in \emph{IEEE Annual Int. Symp.
  on Personal, Indoor, and Mobile Radio Communications (PIMRC)}, Sept 2016, pp.
  1--5.

\bibitem{Polyanskiy_2010}
Y.~Polyanskiy, H.~V. Poor, and S.~Verdu, ``Channel coding rate in the finite
  blocklength regime,'' \emph{IEEE Transactions on Information Theory},
  vol.~56, no.~5, pp. 2307--2359, May 2010.

\bibitem{kleinrockI75_book}
L.~Kleinrock, \emph{Queueing Systems}.\hskip 1em plus 0.5em minus 0.4em\relax
  Wiley, 1975, vol.~I.

\bibitem{Harchol-Balter_book}
M.~Harchol-Balter, \emph{Performance Modeling and Design of Computer Systems:
  Queueing Theory in Action}.\hskip 1em plus 0.5em minus 0.4em\relax Cambridge
  University Press, 2013.

\bibitem{sesia_LTE}
S.~Sesia, I.~Toufik, and M.~Baker, \emph{{LTE} -- The {UMTS} Long Term
  Evolution, From Theory to Practice}.\hskip 1em plus 0.5em minus 0.4em\relax
  John Wiley and Sons, 2009.

\end{thebibliography}

\end{document}